\documentclass[aps,prl,twocolumn,nofootinbib,amsmath,showpacs,superscriptaddress,floatfix]{revtex4}
\usepackage{graphicx,epsfig,longtable}
\usepackage{mathptmx}
\usepackage[light,first]{draftcopy} 
\usepackage{epstopdf}
\usepackage[pdftex]{hyperref}
\usepackage{wasysym}

\newcommand{\beqy}{\begin{eqnarray}}
\newcommand{\eeqy}{\end{eqnarray}}
\newcommand{\bmlet}{\begin{subequations}}
\newcommand{\emlet}{\end{subequations}}
\newcounter{saveeqn}

\def\gsimeq{\,\,\raise0.14em\hbox{$>$}\kern-0.76em\lower0.28em\hbox  
{$\sim$}\,\,}  
\def\lsimeq{\,\,\raise0.14em\hbox{$<$}\kern-0.76em\lower0.28em\hbox  
{$\sim$}\,\,}  

\begin{document}

\title{Observation of large scissors resonance strength in actinides}

\author{M.~Guttormsen}
\email{magne.guttormsen@fys.uio.no}
\affiliation{Department of Physics, University of Oslo, N-0316 Oslo, Norway}
\author{L.A.~Bernstein}
\affiliation{Lawrence Livermore National Laboratory, 7000 East Avenue, Livermore, CA 94550-9234, USA}
\author{A.~B{\"{u}}rger}
\affiliation{Department of Physics, University of Oslo, N-0316 Oslo, Norway}
\author{A.~G{\"{o}}rgen}
\affiliation{Department of Physics, University of Oslo, N-0316 Oslo, Norway}
\author{F.~Gunsing}
\affiliation{CEA Saclay, DSM/Irfu/SPhN, F-91191 Gif-sur-Yvette Cedex, France}
\author{T.W.~Hagen}
\affiliation{Department of Physics, University of Oslo, N-0316 Oslo, Norway}
\author{A.C.~Larsen}
\affiliation{Department of Physics, University of Oslo, N-0316 Oslo, Norway}
\author{T.~Renstr{\o}m}
\affiliation{Department of Physics, University of Oslo, N-0316 Oslo, Norway}
\author{S.~Siem}
\affiliation{Department of Physics, University of Oslo, N-0316 Oslo, Norway}
\author{M.~Wiedeking}
\affiliation{iThemba LABS, P.O. Box 722, 7129 Somerset West, South Africa}
\author{J.N.~Wilson}
\affiliation{Institut de Physique Nucleaire d'Orsay, Bat. 100, 15 rue G. Glemenceau, 91406 Orsay Cedex, France}

\date{\today}

\begin{abstract}

The orbital M1-scissors resonance (SR) has been measured for the first time in the quasi-continuum of actinides. 
Particle-$\gamma$ coincidences are recorded with deuteron and $^3$He induced reactions on $^{232}$Th. 
The residual nuclei $^{231,232,233}$Th and $^{232,233}$Pa show an unexpectedly strong integrated strength 
of $B_{M1} = 11-15 \mu_{n}^{2}$ in the $E_{\gamma}=1.0 - 3.5$~MeV region. 
The increased $\gamma$-decay probability in actinides
   due to the SR is important for cross-section
   calculations for future fuel cycles of fast
   nuclear reactors and may also have impact on
   stellar nucleosynthesis.

\end{abstract}


\maketitle

The $\gamma$ decay of excited atomic nuclei is to a large extend governed by collective 
transitions.
The softest collective M1 mode, the scissors resonance (SR), appears when the deformed proton
and neutron clouds oscillate against each other like the blades of a scissors. 
Such an isovector collective motion was first predicted by Lo Iudice and Palumbo~\cite{iudice1978}.

The radiative strength function (RSF) is a measure of the average electromagnetic properties of 
$\gamma$ transitions in the quasi-continuum region.
The RSF is crucial input for calculating 
neutron-induced reaction cross sections for neutron energies
starting from the keV range. It is particularly important to extrapolate in cases where
measured data are insufficient or lacking, and is relevant
for future and existing nuclear power reactors~\cite{chadwick2011}, and
in stellar nucleosynthesis~\cite{arnould2007,kaeppler2011}.
Experimental constraints on the RSF in this mass region may improve the
predictive power of reaction modelling.

The particular situation with the SR built on the ground state has been
extensively studied in ($\gamma,\gamma^{\prime}$) 
and ($e,e^{\prime}$) reactions. Recently, a review of these experiments 
and various models has been given~\cite{heyde2011}.
The microscopic description of the SR is based on 
cooperative single-particle transitions between orbitals of 
the same angular momentum $\ell$ but different $j=\ell \pm 1/2$.
For deformed rare-earth nuclei one finds experimentally 
integrated strengths of $B_{M1} = 3 - 4\ \mu_N^2$. 
However, the SR is not only built on the 
ground state, but on all states in the nucleus according to the Brink hypothesis~\cite{brink}.
Measurements of the $\gamma$-decay between levels 
in the quasi-continuum show significant higher SR strength.
Here, the two-step cascade method and the Oslo method 
give integrated strengths of $6-7\ \mu_N^2$~\cite{milan2004,schi2006}.

The spins and parities of some SR states in $^{232}$Th and $^{238}$U
have been determined in ($\gamma,\gamma^{\prime}$) and ($e,e^{\prime}$) reactions \cite{heil1988}. 
In addition, ($\gamma,\gamma^{\prime}$) reactions on $^{235,236}$U  
have been reported~\cite{margraf1990,yevetska2010}. 
Again the measured strengths are only
$B_{M1} \sim 3\  \mu_N^2$. Since these experiments rest on the
identification of single states in an energy region housing
$10^{4} - 10^{5}$ levels per MeV, one could expect that not all the 
strength has been resolved as $\gamma$-lines.
    
In this Letter we report for the first time on the observation of the full SR strength 
in the quasi-continuum of actinides nuclei. The data show a clear splitting of the SR strength in $^{233}$Th.
Furthermore, the RSF is found to be independent 
of the excitation energy in the $3-5$~MeV
region, thus supporting the Brink hypothesis~\cite{brink}. 

The Oslo nuclear physics group has developed a method to determine simultaneously 
the nuclear level density and the RSF
from particle-$\gamma$ coincidences~\cite{Schiller00,Lars11}. 
These quantities provide information on the average properties of excited nuclei in the quasi-continuum
and are essential in nuclear reaction theories as they are the only 
quantities needed for a complete description of the $\gamma$ decay at higher excitation energies. 

The experiments were conducted at the Oslo Cyclotron Laboratory (OCL) with a 
12-MeV deuteron and a 24-MeV $^3$He beam bombarding a self-supporting target 
of $^{232}$Th with thickness of 0.968 mg/cm$^2$.    
Particle-$\gamma$ coincidences were measured with the SiRi 
 particle telescope
and the CACTUS $\gamma$-detector systems~\cite{siri,CACTUS}. 
The SiRi detectors were placed in backward direction, 
covering eight angles from $\theta = 126$ to $140^\circ$ 
relative to the beam axis. The front  
and end detectors had a thickness of $130$~$\mu$m and $1550$~$\mu$m, respectively. 
The CACTUS array consists of 28 collimated $5" \times 5"$ NaI(Tl) 
detectors with a total efficiency of $15.2$\% at $E_\gamma = 1.33$~MeV.  

    \begin{figure}[bt]
    \begin{center}
    \includegraphics[clip,width=\columnwidth]{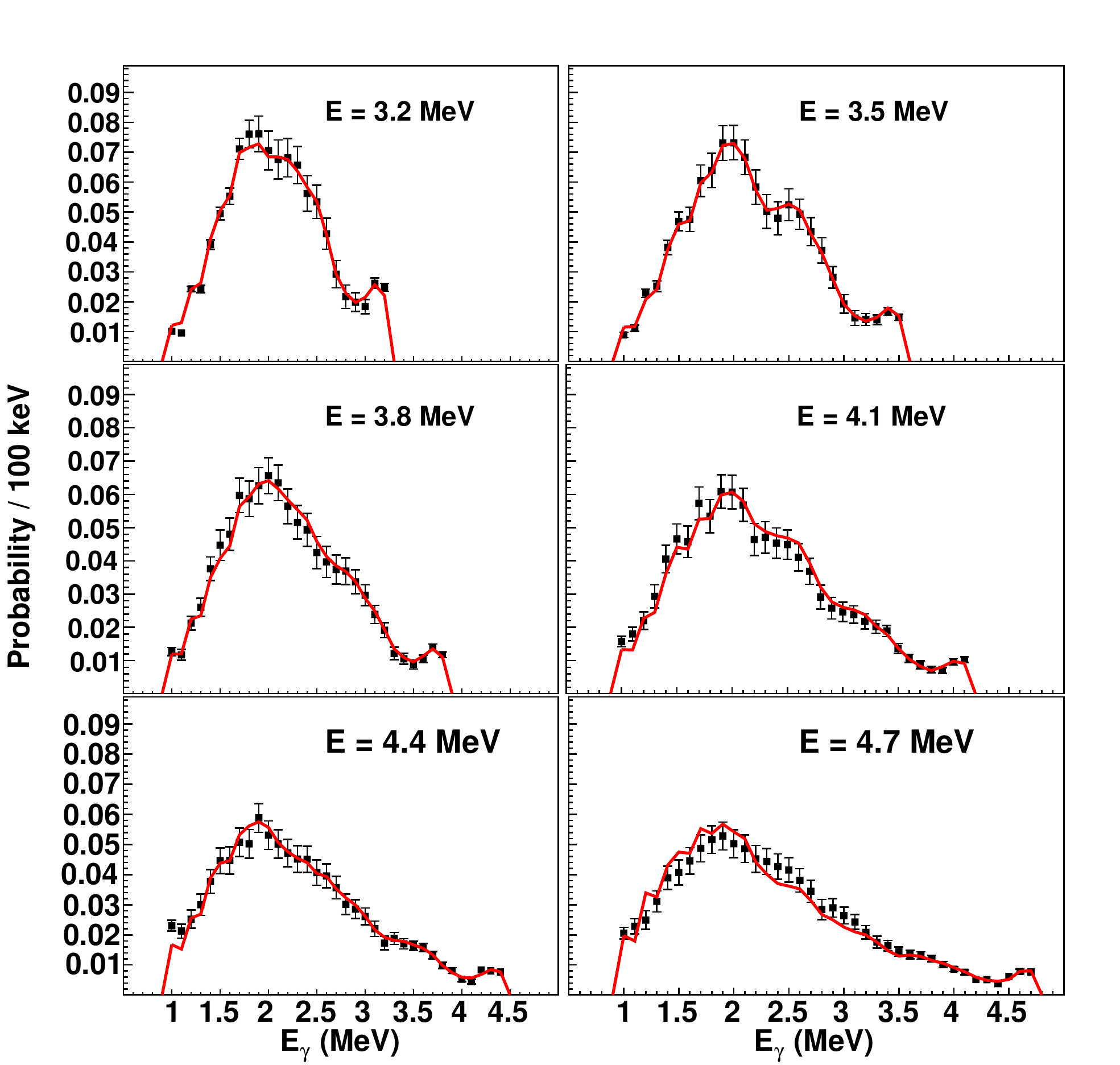}
    \caption{(Color online) First-generation $\gamma$ spectra for $^{233}$Th. The experimental spectra $P$ (squares with error bars) are compared
    with the ones obtained from multiplying the extracted $\cal{T}$ and $\rho$ functions (red line). The initial excitation energy bins $E$ are 100 keV wide.}
    \label{fig:work}
    \end{center}
    \end{figure}
    
The particle-$\gamma$ coincidences with time information are sorted event by event. 
From the known charged-particle type and the kinematics of the reaction, the energies deposited in the telescopes 
can be translated to initial excitation energy $E$ in the residual nucleus. For each energy-bin $E$, 
the $\gamma$-spectra are then unfolded
using the NaI-response function.  
The resulting matrix describes the $\gamma$-ray energy distribution at each bin $E$
and is then the starting point for the Oslo method.

    \begin{table*}[htb]
    \caption{Parameters used for normalization of level density and radiative strength function (see text).} 
    \begin{tabular}{c|c|cccc|cc|c}
    \hline
    \hline
    Nucleus   &$S_n$&   $a$    &$E_1$        &$\sigma(S_n)$&  $D$     &   $\rho(S_n)$  &   $x\rho(S_n)$ &$\langle \Gamma_{\gamma}(S_n)\rangle$\\
              &(MeV)&(MeV$^{-1})$&      (MeV)    &         &  (eV)    &(10$^6$MeV$^{-1}$)&(10$^6$MeV$^{-1}$)&       (meV)                     \\
    \hline
    $^{231}$Th&5.118&   23.91  &  -0.408     &     6.20    & 9.6(15)  & 8.1(16)        &  4.7(8)        &                   26(2)             \\
    $^{232}$Th&6.438&   24.00  &   0.673     &     6.28    &  1.17(35)$^a$&  12.7(38)$^a$   & 6.9(21)$^a$&              33(10)$^a$        \\
    $^{233}$Th&4.786&   24.09  &  -0.389     &     6.13    &16.5(4)   &  4.4(6)        &   2.21(29)     &                   24(2)             \\
    $^{232}$Pa&5.553&   24.00  &  -1.155     &     6.52    &0.53(16)$^a$& 42(12)$^a$   &     20(6)$^a$   &                   33(10)$^a$        \\
    $^{233}$Pa&6.529&   24.09  &  -0.181     &     6.54    &0.42(8)   &  44(10)        &    24(5)       &                   33(10)$^a$        \\
    \hline
    \hline
    \end{tabular}
    \\$^a$) Estimated values from systematics
    \label{tab:parameters}
    \end{table*}

    \begin{figure}[hbt]
    \begin{center}
    \includegraphics[clip,width=\columnwidth]{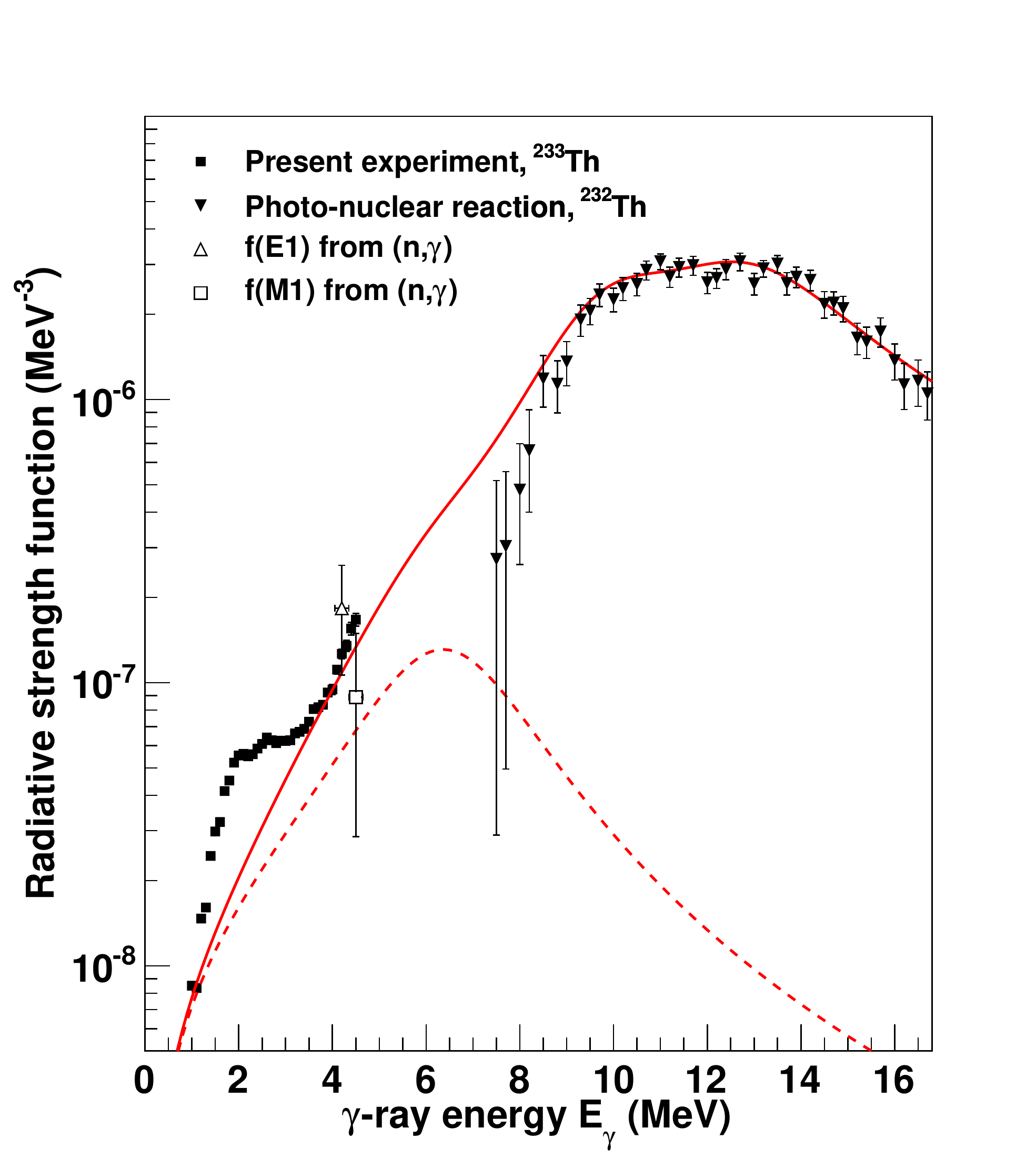}
    \caption{(Color online) Experimental RSF for $^{233}$Th (squares). For comparison, 
    the ($\gamma,{\rm abs}$) reaction on $^{232}$Th~\cite{gurevich} are shown (triangles).
    The solid red line is an estimate of the tail of the $^{232}$Th GEDR (including the GMDR) 
    used for the "background" of the SR.
    Also two data points (open symbols) from the $(n,\gamma)$ reactions are shown~\cite{RIPL3}.
    The resonance parameters ($\omega$, $\Gamma$ and $\sigma$) of the EGLO model~\cite{RIPL3} are 
    (10.9~MeV, 5.66~MeV, 211.4~mb) and (13.87~MeV, 6.68~MeV, 314.3~mb) for the two GEDRs using $T=0$~MeV.
    The GMDR parameters are (6.66~MeV, 4.0~MeV, 9.83~mb).}
    \label{fig:teoexp2}
    \end{center}
    \end{figure}

An iterative subtraction technique has been developed to separate out 
the first-generation (primary) $\gamma$ transitions from the total $\gamma$ cascade~\cite{Gut87}.
It is assumed that the $\gamma$-decay pattern is 
the same whether the levels were initiated directly by the nuclear reaction or by 
decay from higher-lying states. The technique gives the first-generation matrix $P(E,E_{\gamma})$
where $E$ is the initial excitation energy from which the $\gamma$ cascade starts.
In the quasi-continuum, we assume that $P$ is proportional to the 
level density at the final excitation energy $\rho(E-E_{\gamma})$ 
in accordance with Fermi's golden rule~\cite{fermi}. 
Furthermore, the decay is governed by the $\gamma$-transmission 
coefficient ${\cal{T}}(E \rightarrow E-E_{\gamma})$, which according to the Brink 
hypothesis~\cite{brink}, is independent of excitation energy:
\begin{equation}
P(E, E_{\gamma}) \propto   {\cal{T}}(E_{\gamma}) \rho(E-E_{\gamma}).\
\label{eqn:3}
\end{equation}
This allows a simultaneous least $\chi ^2$-fit of the two one-dimensional functions 
${\cal{T}}$ and $\rho$ to the two-dimensional matrix $P$. To test the 
ansatz (\ref{eqn:3}), we have compared
the experimental first-generation spectra and
the ones obtained by multiplying the extracted $\cal{T}$ and $\rho$ functions, see Fig.~\ref{fig:work}. 
The agreement is very good and holds also for the other 11 spectra  
included in the global fit (not shown).
    
The two functions have to be normalized, which means that 
the unknown parameters $A$, $B$ and $\alpha$ in the transformations
\begin{eqnarray}
\tilde{\rho}(E-E_\gamma)&=&A\exp[\alpha(E-E_\gamma)]\,\rho(E-E_\gamma),
\label{eq:array1}\\
\tilde{{\mathcal{T}}}(E_\gamma)&=&B\exp(\alpha E_\gamma){\mathcal{T}} (E_\gamma)
\label{eq:array2}
\end{eqnarray}
must be determined from other experiments.
The parameters $A$ and $\alpha$ can be determined by normalizing $\rho$ to known levels at low 
excitation energy~\cite{ENSDF} and to the level density extracted from 
neutron resonance spacings $D$. We assume a spin distribution~\cite{GC} 
    \begin{equation}
    g(E,I) \simeq \frac{2I+1}{2\sigma^2}\exp\left[-(I+1/2)^2/2\sigma^2\right],
    \label{eq:spindist}
    \end{equation}
where $E$ is excitation energy, $I$ is spin and $\sigma$ is the
spin cut-off parameter. For the actinides studied here, we have typically
$\sigma(S_n) = 6-7 \hbar$, which gives significantly more
high-spin states than populated in the light ion reactions used~\cite{egidy1972}.
Thus, the total 
experimental level density has to be multiplied with a reduction factor to
serve as normalization to the experimental $\rho$ by 
$x\sim \sum_{I=I_{\rm min}}^{I_{\rm max}}g(S_n,I)$, where
$I_{\rm min}$ and $I_{\rm max}$ define the reaction spin window.    

The last parameter $B$ can be determined by reproducing the total $\gamma$-radiation
width $\langle \Gamma_{\gamma} \rangle$ from neutron resonance data. 
In the present work we have followed the 
normalization procedure of \cite{hilde2012} and references therein. 
In the cases where neutron resonance data
are missing, we use values based on the systematics of neighboring nuclei. 
The parameters applied for the normalizations are listed in Table~\ref{tab:parameters}.
The level density parameter $a$ and back-shift parameter $E_1$ are used to
estimate the total level density $\rho$ from the level density spacing $D$
from neutron resonance capture.

Provided that dipole radiation is dominant in the quasi-continuum, 
the RSF can be calculated from the normalized 
transmission coefficient by~\cite{RIPL3}
\begin{equation}
    f (E_{\gamma}) =\frac{1}{2\pi} \frac{ \tilde{{\mathcal{T}}}(E_{\gamma})}{ E_{\gamma}^3}.
    \label{eq:fT}
\end{equation}
Figure~\ref{fig:teoexp2} shows the RSF for $^{233}$Th together with 
the GEDR data~\cite{gurevich} on $^{232}$Th. 

The observed excess in the RSF is interpreted as the SR for several reasons. It is positioned around $E_{\gamma}=2.2$~MeV,
which fits the systematics from nuclei studied in the rare-earth region. 
Also, previous measurements for the SR built on the ground state~\cite{heil1988,margraf1990,yevetska2010} reveal centroids 
around $2.2$~MeV of excitation energy, and several states in these studies are 
proven to be populated by M1 transitions.
To our knowledge, the SR is the only known candidate for a soft collective mode at these energies.

    \begin{figure}[hbt]
    \begin{center}
    \includegraphics[clip,width=\columnwidth]{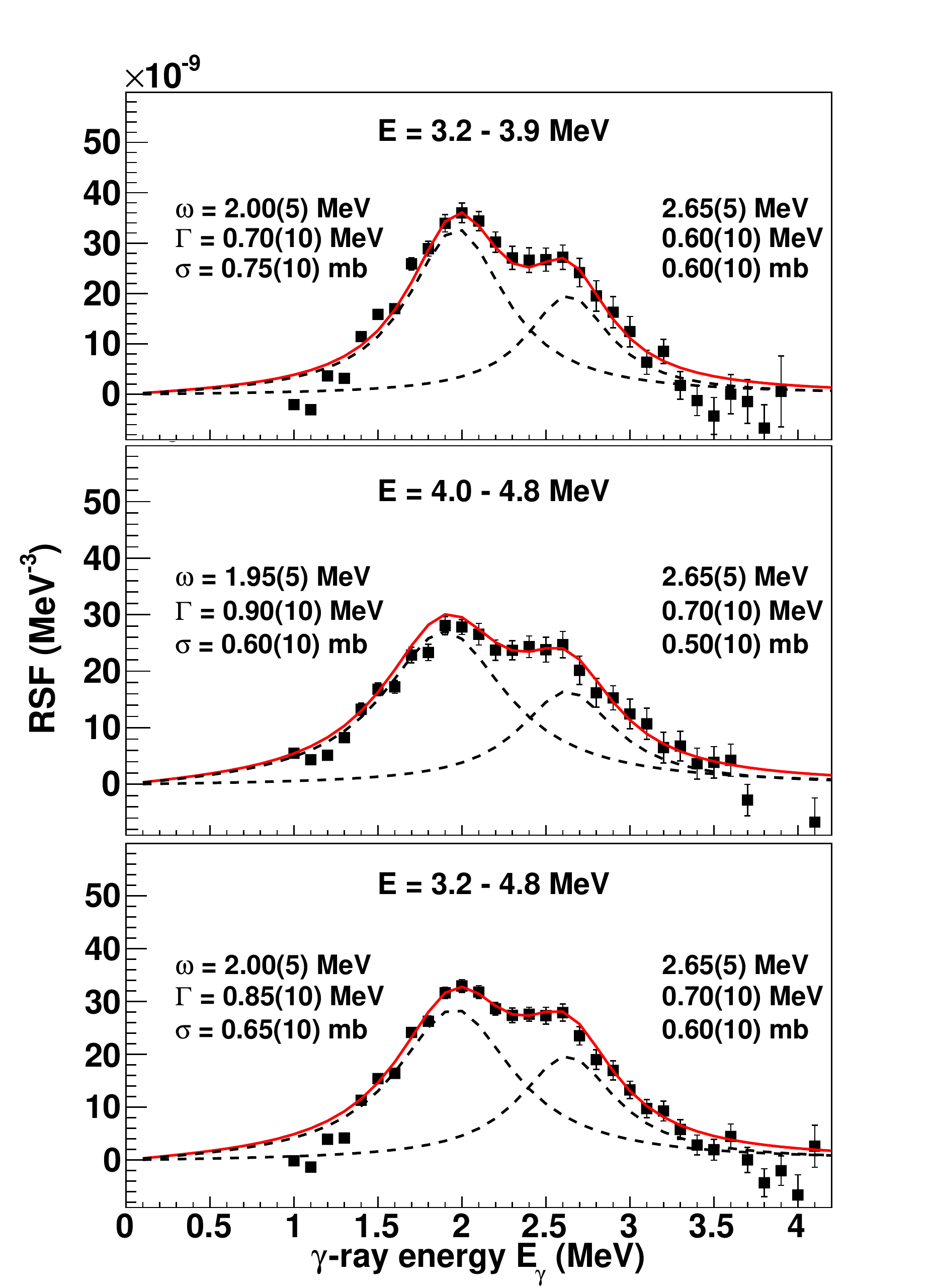}
    \caption{(Color online) The observed SR for various excitation-energy regions of $^{233}$Th. 
    The strengths are obtained by 
    subtracting the underlying tail of the GEDR and GMDR (see Fig.~\ref{fig:teoexp2}).
    The data of the two upper panels are based on statistically independent data sets. 
    The RSF for the lower panel is extracted for simultaneous fitting of the 
    two data sets together, giving approximately the average of the two upper RSFs.
    The resonance centroid $\omega$, width $\Gamma$ and strength $\sigma$ are given for
    the lower and higher resonance components.}
    \label{fig:pygmy3}
    \end{center}
    \end{figure}

In order to extract the SR component of the total RSF, 
we use the extrapolated tail of the giant electric dipole (GEDR)
and giant magnetic dipole (GMDR) resonances, see solid line of Fig.~\ref{fig:teoexp2}. This tail
is tailored to fit the low and high $\gamma$-energy part of the 
experimental RSF data of the various actinides studied. Of course,
this is not an ideal situation, and only photo-nuclear reaction data for $^{232}$Th exist in literature. On
the other hand, the background contributes only $\sim 30$\% to the total RSF. The
uncertainty in $f_{GEDR}$ will introduce some errors in the total SR strength, 
but is negligible for the centroid and the width of the SR energy distribution. 
    
The present $^{232}$Th(d,p)$^{233}$Th experiment gave a rich data set. In fact,
two statistically independent data sets of the $P(E,E_{\gamma})$ matrix could be analyzed and compared.
Figure~\ref{fig:pygmy3} shows the SR energy distributions and resonance parameters of $^{233}$Th 
from initial excitation energies $E=3.2-3.9$~MeV and $E=4.0-4.8$~MeV.
In the lower panel, the result from a fit to the whole excitation region, $E=3.2-4.8$~MeV, is shown for comparison.
The centroids and the strength of the two SRs are almost identical. The width increases somewhat in 
the higher excitation-energy region, but a corresponding reduction in the $\sigma$ parameter maintains the strength.
We conclude that the similarity of the two distributions supports the validity of the Brink hypothesis 
in this energy and mass region.

It is evident from the two data sets that the SR is split into two Lorentzians.
From the resonance parameters of Fig.~\ref{fig:pygmy3}, the integrated $B_{M1}$ 
strengths of the components can be calculated by
\begin{equation}
    B_{M1}=\frac{9\hbar c}{32 \pi ^2}\left( \frac{\sigma \Gamma}{\omega_{M1}}\right),
\end{equation}
giving for the whole excitation region a strength of the first resonance of 
$B_{M1} =9.7(15)\mu_N^2 $ and the second of $5.6(7)\mu_N^2 $. 
The corresponding energy splitting is $\Delta\omega_{M1} = 0.65(3)$~MeV.

\begin{figure}[hbt]
    \begin{center}
    \includegraphics[clip,width=\columnwidth]{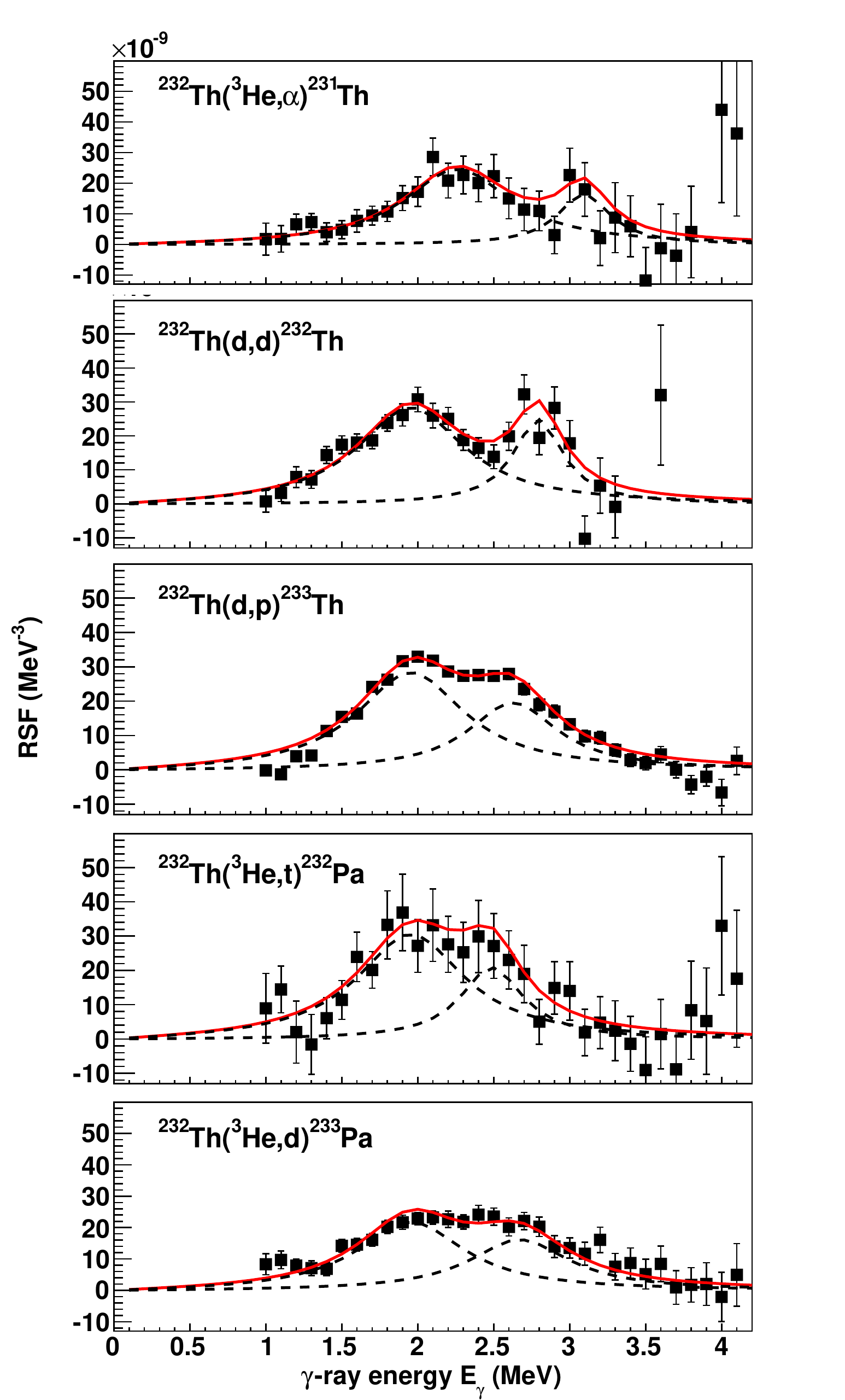}
    \caption{(Color online) The SR for the five nuclei studied. Approximate GEDR tails are
    subtracted from the total RSFs as demonstrated in Fig.~\ref{fig:teoexp2}.}
    \label{fig:pygmy5}
    \end{center}
    \end{figure}

There are various models for the SR properties~\cite{heyde2011}, 
and we choose here the sum-rule approach~\cite{lipparini1989}. 
By multiplying the 
resonance centroid $\omega_{M1}$
with the inverse energy-weighted sum rule $S_{-1}$ we obtain the SR strength:
\begin{equation}
    B_{M1}=\omega_{M1}S_{-1}=\omega_{M1}\frac{3}{16\pi}\Theta_{\rm IV}(g_p - g_n)^2 \mu_N^2.
\end{equation}
We use bare gyromagnetic factors for the protons ($g_p=1$) and neutrons ($g_n=0$).
   Since our measurements are in the quasi-continuum, the isovector moment 
   of inertia $\Theta_{\rm IV}$ is taken as the rigid-body moment of inertia 
    $\Theta_{\rm rigid} =\frac{2}{5}m_N r_0^2 A^{5/3}(1+0.31\delta)$
    with $r_0=1.15$ fm and $\delta$ is the nuclear quadrupole deformation taken from~\cite{moeller1995}.
Figure~\ref{fig:pygmy5} displays the extracted SR energy distributions for the five nuclei measured in this work.
The centroids $\omega_{M1}$ and strengths $B_{M1}$ are summarized in Table~\ref{tab:results}.
The agreement with the predicted sum-rule strength is gratifying.

\begin{table}[]
\caption{Scissors mode parameters (see text).} 
\begin{tabular}{c|c|cc|c}
\hline
\hline
Nuclide&$\delta$&$\omega_{M1}$& $B_{M1}$ & $\omega_{M1} S_{-1}$ \\
          &     &     MeV  &$\mu_N^2$ &   $\mu_N^2$ \\
\hline
$^{231}$Th&0.183&  2.49(20) &  11.2(30)&   17.4      \\
$^{232}$Th&0.192&  2.23(20) &  13.8(40)&   15.8      \\
$^{233}$Th&0.200&  2.24(10) &  15.3(20)&   16.0      \\
$^{232}$Pa&0.192&  2.14(20) &  14.7(40)&   15.1      \\
$^{233}$Pa&0.192&  2.29(20) &  12.7(30)&   16.3      \\
\hline
\hline
\end{tabular}
\\
\label{tab:results}
\end{table}

    Although not completely evident for all five nuclei,
    we assume that the SRs have two components as
    shown as dashed Lorentzians in Fig.~\ref{fig:pygmy5}. Typically, 
    the experimental splitting is $\Delta\omega_{M1} \sim 0.7$~MeV,
    and the ratio of the strengths between the lower and upper resonance components is $B_2/B_1\sim 2$.
    
    The splitting could be due to $\gamma$ deformation, which splits the SR into three components~\cite{lipparini1989} 
    where the two first have
    \begin{eqnarray}
    \omega_1=\left(\cos\gamma +\eta\sin\gamma\right)\omega_{M1},\ B_1=\frac{1}{2}\left(\cos\gamma +\eta\sin\gamma\right)B_{M1},\\
    \omega_2=\left(\cos\gamma -\eta\sin\gamma\right)\omega_{M1},\ B_2=\frac{1}{2}\left(\cos\gamma -\eta\sin\gamma\right)B_{M1},
    \end{eqnarray}
    with $\eta=\sqrt{1/3}$.
    In order to describe the observed splitting of $\sim0.7$~MeV,
    a deformation of $\gamma \sim 15 ^{\circ}$ is required. With this choice, we
    obtain theoretically $B_2/B_1 \sim 0.7$, which is not in accordance with the observed
    ratio. The third SR component is fragmented around $\omega_3= 2\eta\sin\gamma\ \omega_{M1}\sim 0.7$~MeV
    and carries a strength of $B_3\sim 0.5 \mu_N^2$, only. This prediction is very
    difficult (if at all possible) to verify experimentally. 
    
    In conclusion, a large integrated SR strength is found in several actinides with centroids around $\omega_{M1}\sim 2.2$~MeV.
    The strength, which is well described by the inverse-energy weighted sum rule, 
    is about three times larger than the GEDR contribution. 
    
    It is reasonable to believe that the SR will appear for all deformed 
    nuclei in this mass region. The presence of
    the SR will effectively enhance the $\gamma$-decay probability for excitations 
    above the neutron binding energy. 
  As a result, the increased calculated $(n, \gamma)$
  cross sections for actinides with insufficient or
  lacking experimental data could have a
  significant impact on fuel-cycle calculations of
  fast nuclear reactors.
    In addition,
    it has the potential of improving the nuclear-physics aspect of the nucleosynthesis 
    in the actinide region.
    
    The energy splitting of 
    the SR could indicate a deformation
    of $\gamma \sim 15 ^{\circ}$. However, theory predicts that the higher SR component
    has the strongest strength in contradiction with the observations. Therefore, the splitting may be due to other
    mechanisms. 
    
\acknowledgements

We would like to thank E.A.~Olsen, A.~Semchenkov and J.~Wikne at the Oslo Cyclotron Laboratory 
for providing the stable and high-quality deuterium and $^3$He beams during the experiment and 
the Lawrence Livermore National Laboratory for providing the $^{232}$Th target. 
This work was supported by the Research Council of Norway (NFR),
the French national research programme GEDEPEON,  
the US Department of Energy under Contract No. DE-AC52- 07NA27344, and
the National Research Foundation of South Africa.

\vfill
\end{document}